\documentstyle[aps,prb,multicol,psfig]{revtex}

\newcommand{\be}{\begin{equation}}
\newcommand{\ee}{\end{equation}}

\begin{document}


\title{Global versus Local Ferromagnetism in a Model for Diluted Magnetic Semiconductors Studied with Monte Carlo Techniques}

\author{Matthias Mayr$^1$, Gonzalo Alvarez$^2$, and Elbio Dagotto$^2$}

\address{$^1$Max-Planck-Institut f\"ur Festk\"orperforschung, 70569 Stuttgart, Germany}
\address{$^2$National High Magnetic Field Lab and Department of Physics,
Florida State University, Tallahassee, FL 32310, USA}

\maketitle

\begin{abstract}
A model recently introduced for diluted magnetic semiconductors by Berciu and Bhatt (PRL {\bf 87}, 107203 (2001)) is studied 
with a Monte Carlo technique, and the results are compared 
to Hartree-Fock calculations. For doping rates close to the experimentally observed metal-insulator transition, 
a picture dominated by ferromagnetic droplets formed below a $T^\star$ scale emerges. The moments of these droplets 
align as the temperature is lowered below a critical value $T_C$$<$$T^\star$. Our Monte Carlo investigations provide 
critical temperatures considerably smaller than Hartree-Fock predictions. Disorder does not seem to enhance 
ferromagnetism substantially. The inhomogeneous droplet state should be strongly susceptible to changes in doping 
and external fields. 
 
\vskip .3cm
\end{abstract}

\begin{multicols}{2}

Recent advances in experimental techniques have allowed for the introduction of magnetic elements into semiconducting 
hosts beyond the solubility limit. This has stimulated the exciting research area of diluted magnetic semiconductors (DMS) 
with the prospect of 
manipulating the charge as well as the spin degrees of freedom, and its possible technological applications in such novel 
fields as ``spintronics''. Prominent among these 
new compounds is the III-V system Ga$_{1-x}$Mn$_x$As, where ferromagnetic (FM) transition temperatures as high as 110 K were
 obtained.\cite{Ohno_1} Although some of the basic ingredients of the physics of these materials such as the
 local antiferromagnetic (AF) exchange between the Mn spins and the charge carriers are known, the origin of such fairly high critical 
temperatures is still in question. The ferromagnetism itself can be understood as being carrier-induced, in a similar 
fashion 
as the FM state in the double-exchange (DE) model for manganites at intermediate doping.\cite{Dagotto_1} However, alternative 
descriptions based on Ruderman-Kittel-Kasuya-Yosida (RKKY) interactions between the impurity spins are claimed to 
lead to transition 
temperatures comparable to those observed.\cite{Dietl_1} Yet it is not obvious whether RKKY can apply for 
systems with small Fermi energies as is the case for compensated semiconductors. 
Additionally, the effects of disorder 
that accompanies chemical doping was neglected in this approach.\cite{MacDonald_1} Recent annealing studies have shown the relevance of disorder and defects in DMS.\cite{Potashnik_1} It was proposed 
\cite{Berciu_1}, based on Hartree-Fock (HF) calculations, that disorder in the dopant position and hopping 
amplitudes 
has a strong effect on the magnetic properties of these compounds, and should be incorporated in trying to understand 
the behavior of DMS.
  
The above described subtle interplay between localizing and delocalizing tendencies in these models often cannot be 
described correctly in a 
mean-field picture, but it requires nearly exact methods such as Monte Carlo (MC) techniques. A well-known example are the 
manganites in the low-doping regime, 
where mean-field approximations and exact solutions lead to different conclusions about the nature of the metal-insulator 
transition (MIT). The cluster formation and percolative picture that has emerged for these systems cannot be captured properly with mean-field approximations.\cite{Dagotto_1} 
Therefore, here the model introduced in Ref.6 for DMS in the low-doping limit describing electrons in an impurity 
band is investigated using MC methods, and the results will be compared to mean-field calculations. 
  
Although the system under study is actually hole-doped, it is described in  
terms of an electron-doped material. Therefore, the acceptor level is treated as a donor level lying approximately 100 meV 
below the conduction band. 
Furthermore, only the impurity band itself will be considered, whereas the valence (conduction) band 
states are neglected. This 
should be a good approximation as long as the Fermi energy of the ``electrons'' is considerably 
smaller than the gap between the donor state and the lowest conduction level. The simplest such model \cite{Berciu_1} takes into account the hopping between random Mn sites and the AF interaction between the impurity
 spins and the charge carriers. The disorder potential and Coulomb interactions are neglected and the model 
is given by
\begin{eqnarray}
H_{DMS} & = &\sum_{i,j,\alpha} (t_{ij}-\mu\delta_{ij})(c^\dagger_{i\alpha}c_{j\alpha} + 
{\mbox {H.c.}}) \nonumber \\ &  &+ \sum_{i,j,\alpha,\beta}J_{ij}(c^\dagger_{j\alpha}\frac{1}{2}{\bbox{\bf\sigma}}_{\alpha,\beta}c_{j\beta}){\bf
S}_i,
\label{eq:FMKH}
\end{eqnarray}
where $c^\dagger_{i\alpha}$ is the creation operator for an electron at the impurity site $i$ with spin $\alpha$, 
and $t_{ij}$ is the hopping amplitude between sites $i,j$. For two sites at 
distance $r$=$|i-j|$ an exponential form, $t(r)$=$2(1+r/a_B)\exp(-r/a_B)$ Ry, was assumed\cite{Berciu_1}, with $a_B$ the Bohr radius 
associated with the impurity site and Ry its corresponding binding energy $E_b$. ${\bf S}_i$ represents the 
(assumed classical) spin at the Mn ($3d^5$) site with $|{\bf S}_i|$=$5/2$, ${\bbox{\bf \sigma}}$ denotes the vector of Pauli 
matrices, $\mu$ the chemical potential, and $J_{ij}$ describes the non-local interaction between impurity sites and mobile holes. 
Owing to the localized nature of the Mn spins, it can be expressed as $J_{ij}$=$J\exp(-2r/a_B)$, with $J$ an AF 
coupling constant.
$J$ is chosen as the energy unit and periodic boundary conditions (PBC) are imposed.
The effects of an external field can easily be introduced via a Zeeman term.

Following Ref.6, we assume the hole binding energy $E_b$=$112.4$meV (1Ry) and, additionally, we have $J$=15meV.
 It was checked that our conclusions below do not depend crucially on the exact values of the exchange coupling and, 
therefore, the results should be typical for the systems considered. The lattice constant $a$ of the GaAs lattice is 
known to be $5.65$\AA, and $a_B$=7.8\AA.\cite{Shklovskii_1} With these values, the various 
nonlocal couplings $J_{ij}$ and hoppings $t_{ij}$ can easily be calculated once a random distribution of the impurity 
ions within the host lattice (of size $L^3$) has been chosen. In addition, these 
systems are presumably strongly compensated through As antisite donors resulting in a relatively small effective 
hole concentration of 10\% of the Mn ions.\cite{Ohno_1}  

The previous mean-field study of $H_{DMS}$ resulted in some interesting 
conclusions about the magnetic behavior of DMS.\cite{Berciu_1} In particular, an increase in $T_C$ by a factor of $\sim$2
 was reported when a disorder Mn distribution was studied, compared with a periodic distribution. This is a 
challenging counterintuitive prediction that deserves to be tested with 
techniques beyond the HF approach, even more so as it is known that critical temperatures typically are overestimated by 
mean-field methods, as it happens, e.g., in the DE model.\cite{Dagotto_1} Fortunately, Hamiltonian 
(\ref{eq:FMKH}) is especially suited for MC calculations, and the procedure used here has been extensively 
discussed in the study of the manganites.\cite{Dagotto_1}

\begin{figure}
\vspace{-0.3cm}
\centerline{\psfig{figure=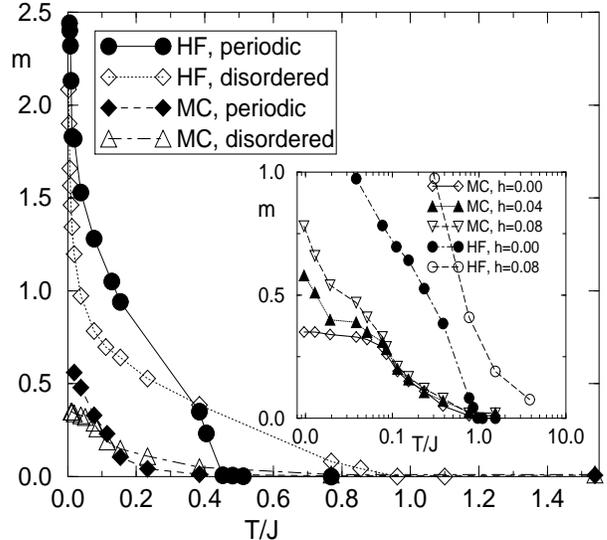,height=8.3cm,angle=0}}
\caption{Magnetization of the Mn spins vs. $T$ for model (\ref{eq:FMKH}) with $x$$\approx$0.02 
and $N$=80 impurity sites. Shown are results of MC as well as HF calculations, for either the random system 
or for impurities in a superlattice (=periodic) configuration. Disorder strongly increases $T_C$ in the mean-field 
description, but much less so in the MC approach. The strong influence of a magnetic field at low temperatures 
is demonstrated in the inset. This is in contrast to HF, where the effect of $h$ is remarkable at all $T$'s (open circles, $h$=0.08).}
\label{Figure_1}
\end{figure}

The MC calculations were performed on a variety of lattice sizes, but the bulk of our work was done studying systems with 
$N$=80 impurity sites and a doping level of $x$$\approx$0.02, 
which lies in the neighborhood of the realistic MIT.\cite{Timm_1} Other lattice sizes ($N$=50-100), with slightly different values 
of $x$, yield similar results and lead to the same conclusions. 
The density $p$ of charge carriers is fixed at $p$=$0.1$ (with respect to the number 
of Mn dopants), unless otherwise noted. The random placement of Mn ions in the GaAs host lattice creates regions 
of high and low impurity density, respectively. Some of the results presented below are for only one such 
configuration of Mn impurities, but it was confirmed that the results for other configurations are within 10$\%$ of the 
data shown. Furthermore, for the configuration selected the HF critical temperature $T^{HF}_C$ is the same as for larger 
systems ($N$=200, 300) with similar values of $x,p$ and $J$. Most 
interesting is the magnetization behavior, with and without applied magnetic field. It is analyzed
 by measuring the magnetization per site, $m$=$M$/$N$ ($M$=$\sqrt{\sum_{ij} {\bf S}_i\cdot{\bf S}_j}$), of the Mn 
spins, neglecting the contribution of the charge carriers. The latter are of much less importance as they have a spin of 
$1/2$ only, and also a density that is only a fraction of that of the Mn spins. Fig.\ref{Figure_1} compares $m$ for the MC as well as for the HF solution, with both results obtained for the {\it same} configuration of Mn 
impurities.\vspace{-0.1cm}
 
\begin{figure}
\centerline{\psfig{figure=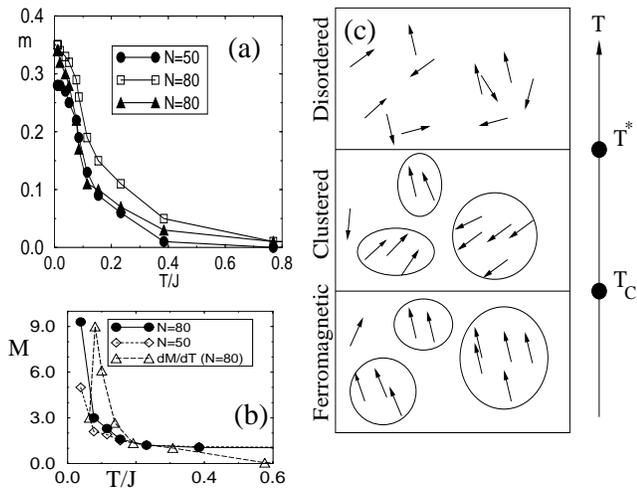,height=6.8cm,angle=0}}
\vspace{0.2cm}
\caption{(a) shows $m$ vs. $T$ for two system sizes, and in the case of the $N$=80 system, the magnetization function 
for a different impurity configuration. In all cases, 0.02$<$$x$$<$0.025. (b) $M$ (scaled by ${|\bf S|}$) for $N$=50, 80
 and d$M$/d$T$ (triangles).
(c) Schematic representation of the development of FM clusters and subsequent alignment as $T$ is lowered. Ordered domains form at 
$T^\star$, and align at the critical point $T_C$.}
\label{Figure_4}
\end{figure}

\begin{figure}
\centerline{\psfig{figure=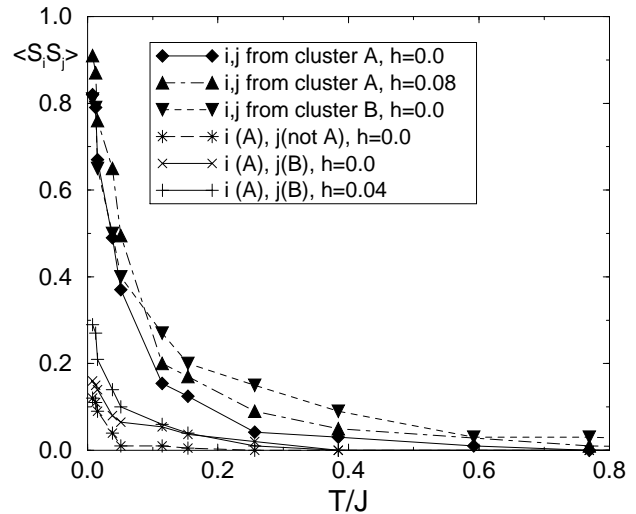,height=7.7cm,angle=0}}
\caption{MC time-averaged spin correlations per site vs. $T$ (scaled by $|{\bf S}|^2$) for the system shown in 
Fig.\ref{Figure_1} for external fields $h$=0.0, 0.04, and 0.08. Correlations are measured either between 
sites i,j of the same cluster (filled symbols, cluster A or B), or between sites that do not belong to the same 
cluster. At high temperatures, the spin-spin correlations are approximately zero at all MC times.}
\label{Figure_2}
\end{figure}

\vspace{-0.1cm}
In this particular case $T^{HF}_C$$\sim$0.85 (similar to earlier results on larger systems \cite{Berciu_1}) 
for the impurity configuration considered, whereas the MC solution suggests an enhancement of FM correlations 
at $\sim$0.3($\pm$0.1). Similar values and comparable deviations are found for other 
configurations and system sizes as well (Fig.\ref{Figure_4}(a)), which supports the notion that mean-field 
treatments generally 
overestimate $T_C$'s. At this same temperature, $M$ (scaled by ${|\bf S|}$) starts to deviate from the totally disordered case $M$=1 as well and also a pronounced change in d$M$/d$T$ is observed (Fig.\ref{Figure_4})(b), signalling the development of FM correlations. Furthermore, the concept of a disorder-enhanced $T_C$ does not appear in agreement with MC. Fig.\ref{Figure_1}
points out the influence of disorder as it also includes HF and MC results for the periodic case. From the 
magnetization data as well as the magnetic 
fluctuation, $\chi$$\propto$$\langle M^2\rangle-\langle M \rangle^2$, a critical behavior around $T^{per}_C$$\sim$0.25$\pm$0.05 can 
be deduced for the periodic system, which is slightly below the critical region for the disordered case. This appears in agreement with 
previous HF results \cite{Berciu_1}, but the reduction in $T_C$ is much less 
pronounced. It will be argued below, however, that the actual transition temperature is lower than $T^{per}_C$. A 
thorough analysis reveals that the disordered system breaks apart into occupied and almost empty domains (in both approaches) 
and the main contribution to $m$ is entirely due to a single, large cluster, with 
average particle density $\langle n \rangle^{cl}$$>$$p$. This leads to an increased temperature where FM correlations start building up compared to the periodic case since the effective local magnetic field is enhanced in this region. Single sites or very small 
clusters are virtually unoccupied and do not order at any finite temperature.  
For this reason the magnetization is never fully saturated in the MC approach as in the HF description, but reaches 
a plateau at about 15\% 
of the possible value for a sample with the given Mn doping rate.  If $x$ is selected considerably lower, it would
 be very unlikely for such a cluster to exist and there should be no finite saturation 
magnetization, as it is observed for samples with $x$$\leq$0.005.\cite{Ohno_1} 
The conclusions above are also supported by analyzing the spin correlations between different clusters A, B
(Fig.\ref{Figure_2}). 
\vspace{-0.3cm}
This reveals that each cluster turns FM as the temperature is reduced\cite{comment_3}, but the overall magnetization will 
remain low 
as the magnetization vectors for different domains point in different directions, whereas in HF they are required to be 
aligned along a chosen $S_z$-axis. In other words, the idea that delocalization of 
charge carriers leads to a spin alignment of the Mn ions across the whole sample\cite{Berciu_1}, appears to be an 
artifact of the approximation employed. It also leads to the unphysical 
observation of totally aligned impurity spins in the extremely diluted limit 
$x$$<$0.001 at very low temperatures. But it is interesting to note from Fig.\ref{Figure_2} that the spin correlations 
between sites of a given cluster and all other sites assume a finite value at temperatures below $\sim$0.08 (10K). 
In addition, 
finite correlations develop between A, B at T$\sim$0.2. The exact value of this temperature
 likely depends on the relative position between two specific domains, but is in good agreement with the value $T_C$=0.08
 quoted above, where all sites were taken into account. Additionally, $M$/${|\bf S|}$ is proportional to the number of spins below this temperature as it should be for a FM ordered state (Fig.\ref{Figure_4}(b)). In any case, 
these values are {\it lower} than $T^{per}_C$ and they suggest two different temperature scales in the problem: 
$T^{\star}$ ($\sim$ 0.3), where isolated magnetic domains are formed, and a smaller true critical temperature 
$T_C$ ($\sim$ 0.1) where long-range order is established. 
Below $T^{\star}$, spins in the vicinity of a given cluster start to align, 
i.e. the magnetic domain grows in size as the gain in kinetic energy outweighs 
the loss in entropy. Further delocalization of the charge carriers then causes the alignment of these domains 
resulting in finite bulk magnetization. The associated $T_C$ is considerably lower than the one 
predicted by mean-field theory.\cite{Kennet_1} This idea is also qualitatively depicted in Fig.\ref{Figure_4}(c). In this scenario 
disorder enhances $T^\star$, but {\it reduces} $T_C$.
\vspace{-0.3cm}
\vspace{-0.1cm}
These percolation-like effects are also visible once a 
magnetic field is applied. As shown in the inset of Fig.\ref{Figure_1}, small fields of $h$$\sim$0.04 (5-10T) are sufficient to 
introduce a robust change in the total magnetization of the sample at low temperatures. 
From Fig.\ref{Figure_2}, it can be concluded that this happens mostly through the mutual alignment of already 
preformed FM areas, 
whereas the magnetization of a given domain remains comparatively unaffected by those small fields. The susceptibilities 
should be even stronger for smaller values of $x$, especially in the insulating regime. A pronounced 
magnetoresistance effect was indeed observed in lightly doped, insulating InAs and GaAs systems in a similar temperature 
regime.\cite{Ohno_PRL} The situation described here is reminiscent of the colossal effects that can be 
observed in transition metal oxides. In that context it was conjectured that disorder by chemical doping leads to 
cluster formation and concomitant large responses to small changes in various parameters.\cite{Burgy_1}\vspace{-0.3cm}
\begin{figure}
\vspace{-0.2cm}
\centerline{\psfig{figure=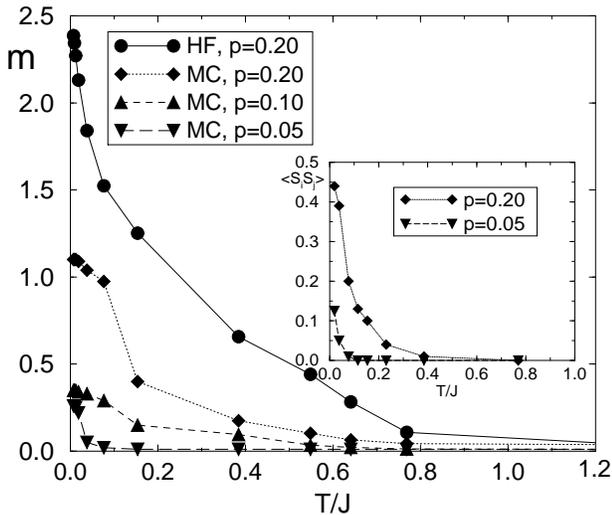,height=7.8cm,angle=0}}
\caption{m vs. $T$ for systems with different doping fractions $p$=0.05, 0.1, and 0.2, and a constant Mn density 
$x$=$0.02$ at zero external field. Weakened compensation leads to an enhanced charge carrier rate, with extended 
FM domains. The inset shows the intercluster (A,B) spin correlations 
for $p$=0.20 and $p$=0.05, demonstrating the influence of $p$ on the correlations.}
\label{Figure_3}
\end{figure}
The possible percolative nature of the magnetic transition in DMS has been discussed recently\cite{Litvinov_1}, and 
it was proposed 
that the relevant analog is the problem of randomly placed spheres which overlap and, thus, form an infinite 
connected area once the radius of the spheres 
is larger than the percolation radius $r_{per}$. In $d$=3, $r_{per}$=1.33$r_o$, with $r_o$ ($\propto$$x^{-1/3}$) the 
interspin distance (and percolation radius) in the periodic case. The MC data ($r_{per}$$\approx$2-3 lattice units) agree 
well with this prediction. In this scenario an increase in $x$ will lead to a smaller $r_{per}$ and, 
therefore, to an increase in $T_C$. At some point, however, $r_{per}$ will be smaller than the typical cluster size at 
$T^{\star}$, which makes $T^{\star}$ the maximum critical temperature possible. This temperature, although slightly lower, 
is in the range of the observed $T^{max}_C$ and small changes in the parameters of $H_{DMS}$ might be sufficient to 
result in higher transition temperatures.  

This can be demonstrated by altering the compensation rate. Fig.\ref{Figure_3} shows $m$ for 
the same impurity configuration as above, but with $p$=$0.20$ and $p$=$0.05$, respectively. As expected, a 
lowered compensation rate leads to enhanced magnetism. The analysis of the spin-correlations (inset Fig.\ref{Figure_3}) 
shows that this is mostly due to an increased 
intercluster spin correlation, whereas the $m$ vs. $T$ curve for a given domain (not shown) is largely unchanged from 
the one at $p$=$0.1$. Thus, the increased number of charge carriers mainly leads to larger droplets, favoring the creation of a coherent FM state. Additionally, the $m$ vs. $T$ curve acquires a more Brillouin-like look.\cite{comment_magnet} The opposite is 
true for the case of even stronger compensation, $p$=$0.05$, where a small reduction in $T_C$ is observed. 

In summary, a model for DMS has been studied with MC methods at low temperatures. For the doping rates considered, the 
system can be described in terms of ferromagnetic droplets. A change in parameters 
such as temperature or external fields can lead to the mutual alignment of these domains, while $T_C$ 
is considerably reduced compared to a mean-field analysis. The HF predictions of a disorder-enhanced 
transition point are not in agreement with nearly exact MC methods. The predicted $T_C$ is somewhat lower than measured 
experimentally, but this value depends considerably on parameters that are not very well known. The inhomogeneous 
state is expected to show large susceptibilities, e.g., a pronounced magnetoresistance effect.

Discussions with M. Berciu, especially about the implementation of the Hartree-Fock method for (\ref{eq:FMKH}), are 
greatfully acknowledged. Part of this work (E.D.) was supported by NSF grant DMR-0122523.\vspace{-0.3cm}  
\bibliographystyle{unsrt}



\begin{thebibliography}{20}
\bibitem{comment_3} $m$ vs. $T$ for a specific domain (see, e.g., Fig.\ref{Figure_2}) can be fitted very well with a Brillouin function.
\bibitem{comment_magnet} Within this family of localized hole models it seems difficult to reproduce the recent annealing 
improved magnetization results \cite{Potashnik_1} which present a rather conventional behavior.
\end{thebibliography}

\newpage





\end{multicols}
\end{document}